\documentclass[11pt]{article}
\usepackage{graphicx} 
\usepackage{amsmath}
\usepackage{amsfonts}   
\usepackage{amssymb}

\usepackage{setspace}
\begin{document}
\date{Today}
\title{{\bf{\Large Voros product, noncommutative inspired Reissner-Nordstr{\"o}m black hole 
and corrected area law }}}

\author{
{\bf {\normalsize Sunandan Gangopadhyay}$^{a,c,}
$\thanks{sunandan.gangopadhyay@gmail.com, sunandan@bose.res.in}},
{\bf {\normalsize Dibakar Roychowdhury}
$^{b,}$\thanks{dibakar@bose.res.in}}\\
$^{a}$ {\normalsize Department of Physics, West Bengal State University, Barasat, India}\\
$^{b}$ {\normalsize  S.N. Bose National Centre for Basic Sciences,}\\{\normalsize JD Block, 
Sector III, Salt Lake, Kolkata 700098, India}\\[0.3cm]
$^{c}${\normalsize Visiting Associate in S.N. Bose National Centre for Basic Sciences,}\\
{\normalsize JD Block, Sector III, Salt Lake, Kolkata 700098, India}\\[0.3cm]
}
\date{}

\maketitle



\begin{abstract}
{\noindent We emphasize the importance of the Voros product 
in defining a noncommutative inspired Reissner-Nordstr\"{o}m black hole. 
The entropy of this black hole is then computed in the tunneling approach and is shown to obey
the area law at the next to leading order in the noncommutative parameter $\theta$.
Modifications to entropy/area law is then obtained by going beyond the semi-classical
approximation. The leading correction to the semiclassical
entropy/area law is found to be logarithmic and its coefficient involves the noncommutative
parameter $\theta$.}
\end{abstract}
\vskip 1cm


\noindent Classical general relativity gives the concept of black hole from which
nothing can escape. This picture was changed dramatically when Hawking
\cite{Hawking1, Hawking2} incorporated the quantum nature into this classical
problem. He showed by combining gravity and quantum mechanics that
black holes emit a spectrum that is similar to a thermal black body
spectrum. This result made the {\it first law of black hole mechanics} \cite{Bardeen} closely analogous to the first law of thermodynamics. This analogy ultimately led to the entropy for black holes and was consistent with the proposal made by Bekenstein \cite{Beken1,Beken2,Beken3,Beken4} that a black hole has an entropy proportional to its horizon area. All the above issues finally led to the famous Bekenstein-Hawking area law for the entropy of black holes given by $S_{BH}=A/4$. 

However, most of these calculations were based on a semiclassical
treatment and also on a commutative spacetime. The standard
Bekenstein-Hawking area law is known to get corrections due to
quantum geometry or back reaction effects \cite{list}.
Recently, modifications to the semiclassical 
area law due to noncommutative (NC) spacetime have also been 
obtained \cite{Modaknon}-\cite{rbsg}. 
The motivation for these investigations 
was that noncommutativity is expected to be relevant at the 
Planck scale where it is known that usual semiclassical 
considerations break down. 

In a recent paper \cite{sgrb}, it has been pointed out that
the Voros star product \cite{Voros},\cite{Lizzi} plays a key role
in the obtention of the mass density of a static, spherically symmetric,
smeared, particle-like gravitational source required in getting the NC
inspired Schwarzschild black hole \cite{gouba},\cite{sunandan}
\footnote{Note that there are ways in which the Moyal star product \cite{sun1} 
is used to incorporate the NC effect in gravity \cite{riv}-\cite{muk}.
The twisted formulation of NC quantum field theory \cite{chaichian}-\cite{sun2} is yet
another way of incorporating the effects of noncommutativity in gravity.}.
Quantum corrections to the semiclassical
Hawking temperature and entropy for NC inspired Schwarzschild black hole
have been obtained next. This is done by first computing the correction 
to the Hawking temperature by going beyond the 
semiclassical approximation in the tunneling method 
\cite{Majhibeyond}-\cite{Tao}. 
Using the corrected form of the Hawking temperature 
and the first law of black hole thermodynamics, 
the entropy is computed. 
The result is seen to contain logarithmic and inverse horizon area 
corrections and holds to $\mathcal{O}(\sqrt{\theta}e^{-M^2/\theta})$. 

In this paper, we carry out the above analysis in the case of NC inspired
Reissner-Nordstr\"{o}m (RN) black hole. To obtain NC effects on the usual
area law, computation of the (NC) Hawking temperature is carried out.
The first law of thermodynamics is then used to obtain the entropy. It is
observed that to the next to leading order in the NC parameter $\theta$, 
in the regime $r_{h}^{2}/(4\theta)>>1$, the (NC) area law is just a NC deformation
of the usual semiclassical area law as in case of the NC inspired Schwarzschild black hole.
Quantum corrections to the area law is then found by following the procedure as mentioned
above for the NC inspired Schwarzschild black hole. 
The coefficient of the logarithmic correction term is explicitly determined 
from the trace anomaly of the stress tensor \cite{Hawking3},\cite{Dewitt}. 
This coefficient is also found to have NC correction. 

To begin the discussion, we follow the analysis in \cite{sgrb} based on the formulational
and interpretational aspects of NC quantum mechanics to highlight the important part played by the Voros star product in writing down the mass and charge densities of a static, spherically symmetric,
smeared, particle-like charged gravitational source required in getting the NC
inspired RN black hole
\begin{eqnarray}
\rho_{\theta}^{(M)}(r)&=&\frac{M}{(4\pi\theta)^{3/2}}
\exp\left(-\frac{r^2}{4\theta}\right)\nonumber\\
\rho_{\theta}^{(Q)}(r)&=&\frac{Q}{(4\pi\theta)^{3/2}}
\exp\left(-\frac{r^2}{4\theta}\right)~.
\label{mass_den}
\end{eqnarray}  
Solving Einstein's equations with the above mass and charge densities incorporated
in the energy-momentum tensor leads to the following 
NC inspired RN metric \cite{Smailrev},\cite{Smail}
\begin{eqnarray}
ds^2 = -f_{\theta}(r) dt^2 + f_{\theta}^{-1}(r)dr^2 + 
r^2(d\tilde\theta^2+\sin^2\tilde\theta d\phi^2) 
\label{1.04}
\end{eqnarray}
where 
\begin{eqnarray}
g_{tt}(r)&=&g^{rr}(r)=f_{\theta}(r)=1-\frac{4M}{r\sqrt\pi}\gamma\left( \frac{3}{2},\frac{r^2}{4\theta}\right)
 + \frac{Q^{2}}{\pi r^{2}}\left[ F(r)+\sqrt{\frac{2}{\theta}}r\gamma\left( \frac{3}{2},\frac{r^2}{4\theta}\right) \right] 
\\
F(r)&=&\gamma^{2}\left( \frac{1}{2},\frac{r^{2}}{4\theta}\right) -\frac{r}{\sqrt{2\theta}}
\gamma\left( \frac{1}{2},\frac{r^{2}}{2\theta}\right)\nonumber.
\label{metric_coef}
\end{eqnarray}

\noindent The event horizon of the black hole can be 
found by setting $g_{tt}(r_h)=0$ in (\ref{1.04}).
Since this equation cannot be solved in a closed form, we take 
the large radius regime ($\frac{r_{h}^2}{4\theta}>>1$) 
where we can expand the incomplete gamma function to 
solve $r_h$ by iteration. Keeping upto the next to leading order 
$\frac{1}{\sqrt{\theta}}e^{-\frac{r^{2}_{0}}{4\theta}}$, we find
\begin{eqnarray}
r_h\simeq r_{0}\left[ 1-\frac{r_{0}}{2\sqrt{\pi\theta}(r_{0}-M)}\left( 2M-\frac{Q^{2}}{\sqrt{2\pi\theta}}\right)
e^{-r^{2}_{0}/(4\theta)} \right]  
\label{1.06}
\end{eqnarray}
where 
\begin{equation}
r_{0}=M+\sqrt{M^{2}-Q^{2}}
\end{equation}
is the horizon radius of the commutative RN black hole.




Now for a general static and spherically symmetric spacetime, 
the Hawking temperature ($T_H$) is related 
to the surface gravity ($\kappa$) by the following relation \cite{Majhi1}
\begin{eqnarray}
T_H=\frac{\hbar\kappa}{2\pi} 
\label{1.061}
\end{eqnarray}
where the surface gravity of the black hole is given by
\begin{eqnarray}
\kappa = \frac{1}{2}\left(\frac{df_{\theta}}{dr}\right)_{r=r_h}.
\label{1.07}
\end{eqnarray}
Therefore in the regime ($\frac{r_{h}^2}{4\theta}>>1$), the Hawking temperature for the 
NC inspired RN black hole is found to be (upto order $\frac{1}{\sqrt{\theta}}e^{-r^{2}_{0}/(4\theta)}$)
\begin{eqnarray}
T_{H}&\simeq&\frac{\hbar}{2\pi r^{3}_{0}}\left[ Mr_{0}-Q^2+\frac{Mr_{0}^2}{\sqrt{\pi\theta}(r_{0}-M)}\left( 3M-r_{0}
-\frac{Q^{2}}{\sqrt{2\pi\theta}}-\frac{(r_{0}-M)}{2\theta}r_{0}^2\right)e^{-r^{2}_{0}/(4\theta)}\right.\nonumber\\  
&&\left. \quad\quad~~~~~+\frac{Q^{2}r_{0}^{4}}{\sqrt{2}4\pi\theta^{2}}e^{-r^{2}_{0}/(4\theta)}\right]~. 
\label{1.10}
\end{eqnarray}
We shall now use the first law of black hole thermodynamics 
to calculate the Bekenstein-Hawking entropy. 
The first law of black hole thermodynamics is given by \cite{Modakex}
\begin{eqnarray}
S=\int\frac{dM}{T_{H}}+\int Y dQ
\label{first_law}
\end{eqnarray}
where 
\begin{eqnarray}
Y&=&-\frac{\Phi_{H}}{T_{H}}-\frac{\partial}{\partial Q}\int\frac{dM}{T_{H}}\\
\Phi_{H}&=&\frac{Q}{r_h}~. \nonumber
\label{y}
\end{eqnarray}
Using eq.(\ref{1.10}), it can be easily shown that $Y=0$ for the NC inspired RN black hole.
Hence the Bekenstein-Hawking entropy upto next to leading order in $\theta$ is found to be 
\begin{eqnarray}
S=\int{\frac{dM}{T_H}}=\pi r^{2}_{0}\left[ 1-\frac{r_{0}}{\sqrt{\pi\theta}(r_{0}-M)}
\left(2M-\frac{Q^{2}}{\sqrt{2\pi\theta}}\right) e^{-r^{2}_{0}/(4\theta)}\right]. 
\label{1.11}
\end{eqnarray}
In order to express the entropy in terms of the 
NC horizon area ($A_{\theta}$), we use eq.(\ref{1.06}) to obtain
\begin{eqnarray}
A_{\theta}= 4\pi r^2_h=4\pi r^{2}_{0}\left[ 1-\frac{r_{0}}{\sqrt{\pi\theta}(r_{0}-M)}
\left( 2M-\frac{Q^{2}}{\sqrt{2\pi\theta}}\right)e^{-r^{2}_{0}/(4\theta)} \right].  
\label{1.12}
\end{eqnarray}
Comparing eq(s)(\ref{1.11}, \ref{1.12}), we find that at the leading order in $\theta$, 
the NC black hole entropy satisfies the area law
\begin{eqnarray}
S=S_{\textrm{BH}}=\frac{A_{\theta}}{4\hbar}~.
\label{1.13}
\end{eqnarray}
This is functionally identical to the Bekenstein-Hawking area law in the commutative space.
     
\noindent Hence we have analytically observed that in the 
regime $\frac{r^2_h}{4\theta}>>1$, the NC version 
of the semiclassical Bekenstein-Hawking area law holds 
upto leading order in $\theta$. This motivates us to investigate 
the corrections to the semiclassical area law upto leading order in $\theta$. 
      
\noindent To do so, we first compute the corrected 
Hawking temperature $\tilde{T}_{H}$. 
Following the tunelling method in \cite{Majhibeyond} which goes beyond
the semiclassical approximation, one obtains the corrected Hawking temperature as
\begin{eqnarray}
\tilde{T}_{H}=T_{H}\left[1+\sum_{i}\frac{\tilde{\beta}_{i}\hbar^{i}}
{(Mr_{h}-Q^{2}/2)^{i}}\right]^{-1}~.
\label{corr_temp}
\end{eqnarray}
Now applying the first law of black hole thermodynamics once again with this corrected Hawking temperature, we obtain the following expression for the corrected entropy/area law :
\begin{eqnarray}
S_{bh}  &=& \frac{A_{\theta}}{4\hbar}+2\pi\tilde{\beta}_{1}\ln A_{\theta} +\mathcal{O}(\sqrt{\theta}e^{-r_{0}^2/(4\theta)})~\nonumber\\      
&=& S_{BH}+2\pi\tilde{\beta}_{1}\ln S_{BH}+\mathcal{O}(\sqrt{\theta}e^{-r_{0}^2/(4\theta)})
\label{corr_entr}
\end{eqnarray}
where $A_{\theta}$ and $S_{BH}$ are defined in (\ref{1.12}) and (\ref{1.13}) respectively. This expression is functionally identical to the corrected entropy/area law for the standard Schwarzschild black hole \cite{Majhitrace,Modakex}. However there is an important difference. This expression of corrected entropy has both noncommutative and quantum corrections. Although here we have restricted ourselves only to the leading order correction due to the NC parameter ($\theta$), one can try to include all order $\theta$ corrections. 
This is technically more involved and we shall not 
address this issue in this paper. 

Now we move on to compute 
the coefficient $\tilde{\beta}_{1}$ in the above expression. 
By making an infinitesimal scale transformation to the metric coefficients in (\ref{1.04}), the coefficient $\tilde{\beta}_{1}$ 
can be related to the trace anomaly in the following way \cite{Modakex}:
\begin{eqnarray}
\nonumber
\tilde{\beta}_{1} &=& -\frac{\sqrt{M^{2}-Q^{2}}}{4\pi\omega}{\textrm{Im}}\int d^{4}x~\sqrt{-g}~\langle{T^{\mu}}_{\mu}\rangle^{(1)}
\nonumber \\
&=& -\frac{(r_{0}-M)}{4\pi\omega}{\textrm{Im}}\int_{r_h}^{\infty}\int_{0}^{-i\beta}\int_{0}^{\pi}\int_{0}^{2\pi} r^2 \sin{\tilde\theta}\langle{T^{\mu}}_{\mu}\rangle^{(1)} drdtd\tilde\theta d\phi
\nonumber \\
&=& \frac{(r_{0}-M)}{2880\pi^{2}\omega}\beta\int_{r_{h}}^{\infty} r^{2}\langle{T^{\mu}}_{\mu}\rangle^{(1)} dr ~.     
\label{coeff_1}
\end{eqnarray}
Here, $\langle{T^{\mu}}_{\mu}\rangle^{(1)}$ is the trace anomaly calculated for the first loop expansion and $\omega$ is given by the Komar energy integral 
\begin{eqnarray}
\omega=\frac{1}{4\pi}\int_{\partial\Sigma}d^{2}x~\sqrt{p^{(2)}}~n^{\mu}\sigma^{\nu}\nabla_{\mu}K_{\nu}
\label{komarint}
\end{eqnarray}
evaluated near the event horizon.
The one loop trace anomaly of the stress tensor for the scalar fields moving in the background of a (3+1) dimensional curved manifold is given by \cite{Hawking3,Dewitt}
\begin{eqnarray}
\langle{T^{\mu}}_{\mu}\rangle^{(1)}=\frac{1}{2880\pi^{2}}
\left(R_{abcd}R^{abcd}-R_{ab}R^{ab}+\nabla_{a}\nabla^{a}R\right)~.
\label{trace_anomaly}
\end{eqnarray}
For the metric (\ref{1.04}), the invariant scalars are explicitly found to be 
\begin{eqnarray}
\nonumber
R_{abcd}R^{abcd} &=& \frac{8}{r^{8}}(7Q^{4}-12MrQ^{2}+6M^{2}r^{2})
\nonumber\\&&-\frac{e^{-r^{2}/(4\theta)}}{4\pi r^{6} \theta^{3}}(\sqrt{2}Q^{2}-4M\sqrt{\pi\theta})\left[ (3Q^{2}-4Mr)r^{4}+4\theta r^{2}(Q^{2}-4Mr)\right] 
\nonumber \\
R_{ab}R^{ab}&=&\frac{4Q^{4}}{r^{8}}+\frac{MQ^{2}e^{-r^{2}/(4\theta)}}{\sqrt{\pi}\theta^{5/2}r^{6}}
(r^{4}-2\theta r^{2})
\nonumber \\
R&=&\frac{e^{-r^{2}/(4\theta)}(Q^{4}-4MQ^{2}\sqrt{2\pi\theta}+8M^{2}\pi\theta)(r^{4}-10r^{2}\theta +8\theta^{2})}{4r^{2}\theta^{3}(\sqrt{2}\pi Q^{2}-4M\pi^{3/2}\sqrt{\theta})}~.
\label{Rab} 
\end{eqnarray}
Note that in the commutative limit ($\theta\rightarrow 0$), the above results match with the known results of the standard vacuum RN spacetime metric \cite{Modakex}, for which $R_{abcd}R^{abcd}=\frac{8}{r^{8}}(7Q^{4}-12MrQ^{2}+6M^{2}r^{2}),~~R_{ab}R^{ab}=\frac{4Q^{4}}{r^{8}}~,~~R=0$. 
To find the trace anomaly (\ref{trace_anomaly}), we now evaluate
\begin{eqnarray}
\nabla_a\nabla^aR &=& \frac{e^{-r^{2}/(4\theta)}}{32\pi r^{6} \theta^{5}}(\sqrt{2}Q^{2}-4M\sqrt{\pi\theta}) \left[ r^{8}(Q^{2}-2Mr+r^{2})-\alpha_{1}+\alpha_{2}\right] ~~~~~\label{nabla}\\
&& {\textrm{where,}}~~~\alpha_{1} = 4r^{6}\theta[5Q^{2}+r(6r-11M)]\nonumber\\
&& ~~~~~~~~~~~\alpha_{2} = 4r^{4}\theta^{2}(9Q^{2}-32Mr+23r^{2})~.\nonumber
\end{eqnarray}
Exploiting all these results, the trace anomaly is computed 
from (\ref{trace_anomaly}), upto the leading order in 
${\cal O}(e^{-r^2/(4\theta)})$, as 
\begin{eqnarray}
\langle{T^{\mu}}_{\mu}\rangle^{(1)}&=&\frac{1}{2880\pi^{2}}
\left[ \frac{52Q^{4}}{r^{8}}-\frac{96MQ^{2}}{r^{7}}
+\frac{48M^{2}}{r^{6}}\right.\nonumber\\
&&\left.+\frac{e^{-r^{2}/(4\theta)}}{32\pi r^{4}\theta^{5}}(\sqrt{2}Q^{2}-4M\sqrt{\pi\theta})
\alpha_{3}-\frac{MQ^{2}e^{-r^{2}/(4\theta)}}{\sqrt{\pi}\theta^{5/2}r^{6}}(r^{4}-2\theta r^{2})\right] 
\label{trace_anomaly_1}
\end{eqnarray}
where,
\begin{eqnarray}
\alpha_{3}&=&r^{6}(Q^{2}-2Mr+r^{2})-4r^{4}\theta(5Q^{2}+6r^{2}-11Mr)    \nonumber\\
&&~~~~~~~~~~~~~~~+4\theta^{2}r^{2}(3Q^{2}-24Mr+23r^{2})-32\theta^{3}(Q^{2}-4Mr)~.
\label{z42}
\end{eqnarray}
Substituting this in (\ref{coeff_1}) and performing the integral yields
\begin{eqnarray}
\tilde{\beta}_{1}&=&\frac{(r_0 -M)}{2880\pi^{2}\omega}\beta \left[ \frac{52Q^{4}}{5r_{h}^{5}}-\frac{24MQ^{2}}{r_{h}^{4}}+\frac{16M^{2}}{r_{h}^{3}}+\frac{2MQ^{2}e^{-r^{2}_{0}/(4\theta)}}{\theta\sqrt{\pi\theta}r^{3}_{0}}(2\theta -r^{2}_{0})\right.\nonumber\\
&&\left.+\frac{e^{-r^{2}_{0}/(4\theta)}}{32\pi}(\sqrt{2}Q^{2}-4M\sqrt{\pi\theta})\alpha_{4}\right] 
\label{coeff_1a}
\end{eqnarray}
where ,
\begin{eqnarray}
\alpha_{4}=\frac{2Q^{2}}{\theta^{2}}\left( \frac{r^{3}_{0}}{\theta^{2}}-\frac{16}{r_{0}}-\frac{14r_{0}}{\theta}-\frac{32\theta}{r^{3}_{0}}\right) +\left( \frac{2r^{4}_{0}}{\theta^{4}}-\frac{28r^{2}_{0}}{\theta^{3}}\right)
(r_{0}-2M)+\frac{16}{\theta^{2}}(r_{0}+2M)+\frac{256M}{\theta r^{2}_{0}}~. 
\end{eqnarray}
To compute $\omega$, we calculate the Komar energy 
integral (\ref{komarint}). For the spacetime metric (\ref{1.04}) one has
the following expressions for the Killing vectors ($K^\mu$), its inverse
($K_\nu$) and the unit normal vectors ($n^\mu$, $\sigma^\nu$)
\begin{eqnarray}
&& K^{\mu} = (1,0,0,0),~~~K_{\nu} = -f_{\theta}(1,0,0,0)\\
&& n^{\mu} = f_{\theta}^{-1/2}(1,0,0,0)\\
&& \sigma^{\nu} = f_{\theta}^{1/2}(0,1,0,0)\\
&& \sqrt{p^{(2)}} = r^2\sin\theta.
\label{kom}
\end{eqnarray}
Using these, eq.(\ref{komarint}) is simplified as
\begin{eqnarray}
\omega=\frac{1}{8\pi}\int_{\tilde\theta=0}^{\pi}\int_{\phi=0}^{2\pi}
r^2\sin\tilde\theta~(\partial_{r}f_{\theta})~d\tilde\theta d\phi~.
\label{komarsimp}
\end{eqnarray}
Finally, integrating over the angular variables,  
near the event horizon (\ref{1.06}), the above expression for $\omega$ simplifies to
\begin{eqnarray}
\omega&=&(r_{0}-M)\left[ 1-\frac{Mr_{0}}{\sqrt{\pi\theta}(r_{0}-M)}\left( 1+\frac{r^{2}_{0}}{2\theta}\right)e^{-r^{2}_{0}/(4\theta)}\right.\nonumber\\
&&\left.+\frac{Q^{2}r^{3}_{0}}{\sqrt{2}4\pi\theta^{2}(r_{0}-M)}
e^{-r^{2}_{0}/(4\theta)}+\frac{Q^{2}}{\sqrt{\pi\theta}(r_{0}-M)^{2}}\left( 2M-\frac{Q^{2}}{\sqrt{2\pi\theta}}\right)e^{-r^{2}_{0}/(4\theta)}\right]~. 
\label{komar1b}
\end{eqnarray}
Substituting this in the expression for $\tilde{\beta}_{1}$ 
in eq.(\ref{coeff_1a}), we obtain :
\begin{equation}
\tilde{\beta}_{1}=\frac{r^{2}_{0}}{1440\pi(r_{0}-M)}\left[ \alpha_{5}+\left( \frac{52Q^{4}}{5r^{5}_{0}}-\frac{24MQ^{2}}{r^{4}_{0}}+\frac{16M^{2}}{r^{3}_{0}}\right) \alpha_{6}\right] 
\label{coeff_1b}
\end{equation}
where,
\begin{eqnarray}
\alpha_{5}&=&\frac{52Q^{4}}{5r^{5}_{h}}-\frac{24MQ^{2}}{r^{4}_{h}}+\frac{16M^{2}}{r^{3}_{h}}+\frac{2MQ^{2}}{\theta\sqrt{\pi\theta}r^{3}_{0}}(2\theta -r^{2}_{0})
e^{-r^{2}_{0}/(4\theta)}
\nonumber \\
&&+\frac{1}{32\pi}(\sqrt{2}Q^{2}-4M\sqrt{\pi\theta})e^{-r^{2}_{0}/(4\theta)}\left[\frac{2Q^{2}}{\theta^{2}}\left( \frac{r^{3}_{0}}{\theta^{2}}-\frac{16}{r_{0}}-\frac{14r_{0}}{\theta}-\frac{32\theta}{r^{3}_{0}}\right) 
\right.\nonumber \\ 
&&\left.+\left( \frac{2r^{4}_{0}}{\theta^{4}}-\frac{28r^{2}_{0}}{\theta^{3}}	\right) (r_{0}-2M)+\frac{16}{\theta^{2}}(r_{0}+2M)+\frac{256M}{\theta r^{2}_{0}}\right]\nonumber\\
\alpha_{6}&=&\frac{Mr_{0}}{\sqrt{\pi\theta}(r_{0}-M)}\left( 1+\frac{r^{2}_{0}}{2\theta}\right)e^{-r^{2}_{0}/(4\theta)}
-\frac{Q^{2}r^{3}_{0}}{\sqrt{2}4\pi\theta^{2}(r_{0}-M)}e^{-r^{2}_{0}/(4\theta)}\nonumber\\
&&-\frac{Q^{2}}{\sqrt{\pi\theta}(r_{0}-M)^{2}}\left(2M-\frac{Q^{2}}{\sqrt{2\pi\theta}}\right)e^{-r^{2}_{0}/(4\theta)}~.
\label{5a}
\end{eqnarray}
Finally using eq(s)(\ref{corr_entr}) and (\ref{coeff_1b}), 
we find the cherished result for the corrected entropy/area law (upto leading order in $\theta$) for the NC inspired RN black hole
\begin{eqnarray}
S_{bh}  &=& \frac{A_{\theta}}{4\hbar}+\frac{r^{2}_{0}}{720(r_{0}-M)}\left[ \alpha_{5}+\left( \frac{52Q^{4}}{5r^{5}_{0}}-\frac{24MQ^{2}}{r^{4}_{0}}+\frac{16M^{2}}{r^{3}_{0}}\right) \alpha_{6}\right] \ln\frac{A_{\theta}}{\hbar} + \mathcal{O}(\sqrt{\theta}e^{-\frac{M^2}{\theta}})
\nonumber\\        
 &=& S_{BH}+ \frac{r^{2}_{0}}{720(r_{0}-M)}\left[ \alpha_{5}+\left( \frac{52Q^{4}}{5r^{5}_{0}}-\frac{24MQ^{2}}{r^{4}_{0}}+\frac{16M^{2}}{r^{3}_{0}}\right) \alpha_{6}\right]\ln S_{BH}+\mathcal{O}(\sqrt{\theta}e^{-\frac{M^2}{\theta}}).\nonumber\\
\label{corr_entr2}
\end{eqnarray}
This is the general expression for the entropy of NC 
inspired RN black hole where both the NC and quantum 
effects have been taken into account. 
The first term in this expression is the semiclassical 
entropy and the next term is the leading correction. 
It is logarithmic in nature. The coefficient of the logarithmic correction is different from the standard RN black hole \cite{Majhitrace, Modakex} due to the presence of noncommutative parameter ($\theta$). In the commutative limit $\theta\rightarrow 0$, the expression for the corrected entropy exactly matches with the standard RN case where the coefficient of the leading correction is $\frac{1}{90}$, obtained in the path integral \cite{Hawking3}, euclidean \cite{fur} and tunneling \cite{Majhitrace},\cite{Modakex} formalisms.
Also, in the $Q\rightarrow0$ limit, eq.(\ref{corr_entr2}) reduces to the entropy of the NC inspired Schwarzschild black hole \cite{sgrb}.

We conclude by making the following comments. 
In this paper we have once again emphasized the importance of the 
Voros star product in writing down the mass and charge densities 
of a NC inspired RN black hole. 
To point out the role played by the Voros product, 
we need to take recourse to a rigorous formulation 
of NC quantum mechanics \cite{gouba},\cite{sunandan}
as in our earlier paper \cite{sgrb}. 

We have then studied the status of the entropy/area law 
along with corrections of the NC inspired RN black hole. 
A general result for this black hole 
entropy/area law was found, taking both quantum 
and NC effects into account. For this we first 
used the tunneling method by going beyond the 
semiclassical approximation and calculated the 
corrected Hawking temperature. 
Using this modified temperature and the 
first law of black hole thermodynamics we then calculated 
the corrected entropy. The (NC) semiclassical 
Bekenstein-Hawking value was reproduced at the 
next to leading order in $\theta$ and higher order correction contained 
logarithm of horizon area. 
The coefficient of the logarithmic term was fixed by using the trace anomaly of the scalar 
field stress tensor. The trace anomaly and the 
Komar energy integral for the NC inspired RN  
metric were explicitly calculated to determine 
this coefficient. The value of the 
coefficient was found to have NC correction. 
We also show that the commutative limit of 
the corrected entropy/area law of this black 
hole matches with the standard result 
for the RN black hole \cite{Hawking3,fur,Majhitrace}.

\section*{Acknowledgements} One of the authors DRC thanks the Council
of Scientific and Industrial Research (CSIR), Government of India,
for financial support. SG would like to thank the S.N. Bose National
Centre for Basic Sciences where a considerable part of the work was completed.

\end{document}